# LONGITUDINAL BEAM DYNAMICS DESIGN FOR SUPER TAU-CHARM FACILITY

Linhao Zhang[*], Tao Liu, Sangya Li, Jingyu Tang[†], and Qing Luo[‡]
University of Science and Technology of China, 230026, Hefei, Anhui, China

*Abstract*

The project of Super Tau-Charm Facility (STCF) proposed in China, as a new-generation high-luminosity e+/e- collider in the low-energy region with the center-of-mass energy of 2—7 GeV, is well underway. The luminosity is targeted at $1.0\times10^{35}$ cm$^{-2}$s$^{-1}$ at the optimized beam energy of 2 GeV. Longitudinal beam dynamics becomes of great importance for the STCF due to the constraints from the novel beam-beam effect called coherent X-Z instability and severe beam collective effects. In this paper, we will develop an iterative optimization model for the STCF longitudinal beam dynamics design, which takes into account the influence of transverse dynamics, coherent X-Z instability, and collective effects.

## INTRODUCTION

The STCF proposed in China is a new-generation super high luminosity e+/e- collider in the low-energy region spanning the center-of-mass energy of 2—7 GeV. It aims to explore the rich physics in the tau-charm energy range and even search for new physics beyond the standard model [1]. The goal luminosity of STCF reaches $1.0\times10^{35}$ cm$^{-2}$s$^{-1}$ at the optimized beam energy of 2 GeV, which is two orders of magnitude higher than that of the existing e+/e- collider in the tau-charm field in China, BEPCII.

To achieve such high luminosity, a large Piwinski angle combined with the crab-waist collision scheme has been widely recognized as an effective approach [2], and adopted in the new-generation e+/e- colliders such as SuperKEKB [3], BINP-SCTF [4], FCC-ee [5], and CEPC [6], etc. On one hand, through introducing a large Piwinski angle $\phi$, the vertical beta function at the interaction point (IP) $\beta_y^*$ can be squeezed into the level of effective bunch length $\sigma_z/\phi$, to significantly raise the luminosity; On the other hand, the synchro-betatron coupling resonance introduced by the large Piwinski angle can be suppressed by the crab-waist correction scheme using crab sextupoles properly positioned on both sides of the IP. However, this scheme will introduce a novel beam-beam effect called coherent X-Z instability [7], which imposes stringent constraints on the longitudinal beam dynamics by requiring the horizontal beam-beam parameter $\xi_x$ to be much less than the synchrotron tune $\nu_z$.

Furthermore, the STCF is characterized by the beam properties of low energy, small emittance, high bunch intensity and large bunch numbers, which means that the STCF faces significant beam collective effects. This also places strict limitations on the longitudinal parameters such as bunch length and energy spread.

Therefore, special attention is paid to the longitudinal beam dynamics design for STCF in this paper, which requires iterative optimization with transverse dynamics, beam-beam effects, and collective effects, etc., in order to search for possible optimal solutions.

## DESIGN CONSIDERATIONS

The following are the specific considerations and requirements.

### Luminosity

For fully symmetric flat electron-positron beam collisions, the relationship between luminosity and the associated parameters is illustrated in Fig. 1. One can see that the luminosity $L$ is closely related to total beam current $I$, vertical beta function at IP $\beta_y^*$, vertical beam-beam parameter $\xi_y$, and hourglass factor $F_h$ that is always less than 1. To achieve the goal luminosity of $1.0\times10^{35}$ cm$^{-2}$s$^{-1}$ at the optimized beam energy of 2 GeV, the beam current of 2 A and submillimeter $\beta_y^*$ of 0.6 mm are first identified, which suggests that $\xi_y$ is at least larger than 0.07 assuming $F_h$ to be 1. This also implies a limitation on the bunch length $\sigma_z$ since $\xi_y$ is inversely proportional to $\sigma_z$ under a large Piwinski angle $\phi$ with the total crossing angle of $2\theta = 60$ mrad. Additionally, it is noted that the time resolution at the detector requires $\sigma_z$ not larger than 12 mm.

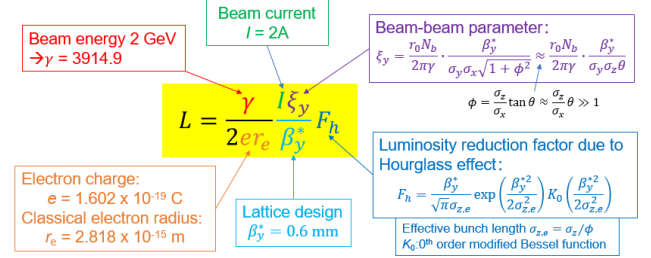

Figure 1: Luminosity and the correlated parameters.

### Coherent X-Z instability

Coherent X-Z instability, as a newly discovered coherent beam-beam interaction under a large Piwinski angle, primarily leads to an increase in the horizontal emittance $\varepsilon_x$ [8]. Considering the coupling between horizontal and vertical emittances, it eventually results in an increase in the vertical emittance $\varepsilon_y$ and thus a collapse of the luminosity. This instability cannot be suppressed through beam feedback systems but can only be avoided through appropriate parameter optimization. Typically, a stringent requirement of $\xi_x \ll \nu_z$ is imposed to prevent its occurrence. In the case of STCF, $\nu_z$ is at least 3 times larger than $\xi_x$.

---

[*] zhanglinhao@ustc.edu.cn
[†] jytang@ustc.edu.cn
[‡] luoqing@ustc.edu.cn

*Lattice design and damping wigglers*

Lattice design is crucial for achieving high luminosity by ensuring the required optical parameters at IP and by optimizing nonlinear dynamics aperture. Additionally, it defines the momentum compaction factor $\alpha_p$ and natural energy spread $\sigma_\delta$ in the electron storage ring through five synchrotron radiation integrals. These parameters ($\alpha_p$ and $\sigma_\delta$) are of utmost importance in the longitudinal dynamics.

Particularly, the STCF needs to introduce damping wigglers to control the damping time and maintain beam emittance almost constant throughout the entire energy range. This inevitably increases the synchrotron radiation energy loss per turn, raising the demand for RF power, and also increases the natural energy spread, resulting in a proportional growth in bunch length. The lattice including damping wigglers has been designed for STCF [9].

*Non-impedance-induced collective effects*

Non-impedance-induced collective effects mainly include the intrabeam scattering (IBS) and Touschek effect. IBS typically refers to multiple small-angle Coulomb scattering processes that do not immediately cause particle loss in the bunch but increase the equilibrium bunch length, energy spread and transverse emittances. On the other hand, the Touschek effect refers to single large-angle scattering processes that limit the lifetime of the stored beam (i.e., the Touschek lifetime). These two effects are directly related to the 6D phase-space size and thus associated with longitudinal parameters such as the bunch length. Bunch lengthening caused by a higher harmonic RF system and longitudinal impedance can help mitigate the IBS effect and improve the Touschek lifetime to some extent. In our case, the ELEGANT code [10] is used to calculate IBS and the Touschek lifetime.

*Impedance-induced collective effects*

Impedance-induced collective effects on a single bunch are also crucial for determining the equilibrium bunch parameters. These effects include bunch lengthening due to potential well distortion (PWD), longitudinal microwave instability (LMWI), and transverse mode coupling instability (TMCI), as expressed below respectively:

$$\left(\frac{\sigma_z}{\sigma_{z0}}\right)^3 - \frac{\sigma_z}{\sigma_{z0}} = \frac{I_b \alpha_p}{4\sqrt{\pi}\nu_z^2 E/e}\left(\frac{R}{\sigma_{z0}}\right)^3 \mathrm{Im}\left(\frac{Z_{//}}{n}\right)_{\mathrm{eff}}, \quad (1)$$

$$I_b^{\mathrm{LMWI}} = \frac{\sigma_z}{R}\frac{\sqrt{2\pi}\alpha_p E/e}{|Z_{//}/n|_{\mathrm{eff}}}\sigma_\delta^2, \quad (2)$$

$$I_b^{\mathrm{TMCI}} = \frac{\sigma_z}{R}\frac{4\sqrt{\pi}\nu_z E/e}{\langle\beta\rangle\mathrm{Im}Z_\perp^{\mathrm{eff}}}, \quad (3)$$

where $\sigma_{z0}$, $R$, $I_b$, $\langle\beta\rangle$, $\mathrm{Im}(Z_{//}/n)_{\mathrm{eff}}$, $\mathrm{Im}Z_\perp^{\mathrm{eff}}$ are the natural bunch length at zero beam current, the average radius of ring, single-bunch current, the average beta function of ring, the longitudinal normalized effective impedance and transverse effective impedance, respectively.

These three collective effects are all related to longitudinal parameters like the synchrotron tune and bunch length. Therefore, it is necessary to design the longitudinal parameters for STCF appropriately in order to control the bunch lengthening caused by PWD within 12 mm and to prevent the occurrence of LMWI and TMCI.

## LONGITUDINAL PARAMETER DESIGN

The above considerations can be divided into two categories: one is the input factors, including the transverse lattice design parameters (considering damping wigglers), requirements for coherent X-Z instability, and IBS; the other is the design optimization objectives, including the satisfaction of bunch length for luminosity and detector requirements (Goal 1), no occurrence of impedance-induced single-bunch collective instabilities (Goal 2), and a minimum Touschek lifetime of 300 s (Goal 3). Figure 2 illustrates the impact of these factors on longitudinal dynamics parameters and their interrelationships.

The design process is as follows: 1) Given the five synchrotron radiation integrals obtained from lattice design that includes damping wigglers, calculate the beam energy spread under the balance of radiation damping and quantum excitation; 2) Taking into account the requirements for coherent X-Z instability, compute the bunch length and synchrotron tune without considering IBS effects; 3) Incorporate radiation damping, quantum excitation, and IBS to obtain the new equilibrium energy spread and bunch length, and check if they meet Goal 1 and Goal 2; 4) Calculate the required RF voltage for achieving the new equilibrium bunch length and its corresponding RF energy acceptance (which should be greater than the momentum aperture from transverse dynamics tracking), and then evaluate the Touschek lifetime to check if Goal 3 is fulfilled.

To achieve the optimization objectives, adjustments may be required in the input constraint parameters. This can involve modifying the synchrotron tune, bunch length, and potentially lattice and damping wiggler parameters, to continue iterative optimization. It is important to prioritize beam stability (Goal 2) and achieving the luminosity target (Goal 1) when the three objectives cannot be simultaneously satisfied. Notably, the Touschek lifetime relies heavily on the optimization of transverse nonlinear dynamics, which is considered highly challenging for STCF[9].

After preliminary optimization, the longitudinal dynamics parameters for STCF have been initially determined. The bunch length is 8.04/8.94 mm (without/with IBS), the energy spread is 7.88/8.77×10⁻⁴ (without/with IBS), and the synchrotron tune is 0.0099. Correspondingly, the required RF voltage is 1.2 MV, providing the RF energy acceptance of 1.56%. Considering IBS, the luminosity can reach $1.45\times10^{35}$ cm$^{-2}$s$^{-1}$ with $\xi_y$ = 0.111. Table 1 shows the main parameters of STCF at 2 GeV, along with a comparison to the main parameters of SuperKEKB LER.

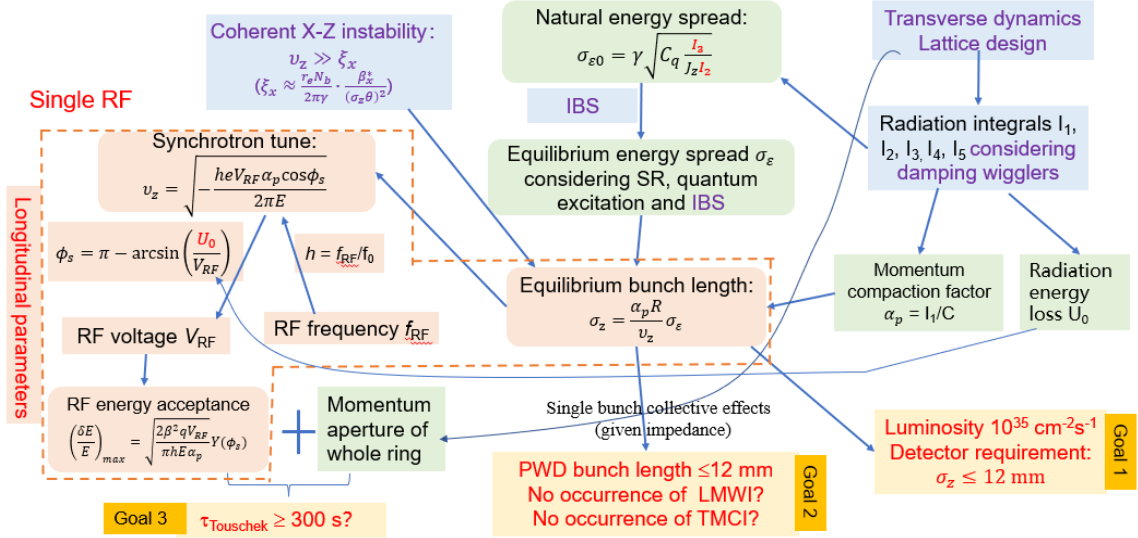

Figure 2: The interrelationships between different limiting factors and longitudinal dynamics parameters.

Table 1: Main Parameters of STCF in Comparison with SuperKEKB LER. Values in Parentheses Denote Parameters without IBS

| Parameters | STCF | SuperKEKB LER |
|---|---|---|
| $E$ [GeV] | 2 | 4 |
| $C$ [m] | 616.76 | 3016.315 |
| $2\theta$ [mrad] | 60 | 83 |
| $\varepsilon_x$ [nm] | 4.47 (3.12) | 3.2 (1.9) |
| $\beta_x^*/\beta_y^*$ [mm] | 40/0.6 | 32/0.27 |
| $\nu_x/\nu_y$ | 31.552/24.572 | 44.530/46.570 |
| $\alpha_p$ [$10^{-4}$] | 10.27 | 3.25 |
| $\sigma_\delta$ [$10^{-4}$] | 8.77 (7.88) | 8.14 (7.96) |
| $I$ [A] | 2 | 3.6 |
| $n_b$ | 512 | 2500 |
| $I_b$ [mA] | 3.91 | 1.44 |
| $U_0$ [keV] | 273 | 1870 |
| $\tau_x/\tau_z$ [ms] | 30.1/15.0 | 43.2/21.6 |
| $f_{RF}$ [MHz] | 497.5 | 508.9 |
| $h$ | 1024 | 5120 |
| $V_{RF}$ [MV] | 1.2 | 9.4 |
| $\nu_z$ | 0.0099 | 0.0247 |
| $\sigma_z$ [mm] | 8.94 (8.04) | 6.0 (5.0) |
| $\delta_{RF}$ [%] | 1.56 | 2.52 |
| $\xi_x/\xi_y$ | 0.0032/0.111 | 0.0028/0.083 |
| $L$ [cm$^{-2}$s$^{-1}$] | $1.45 \times 10^{35}$ | $8 \times 10^{35}$ |

## EVALUATIONS AND DISCUSSIONS

### Impact on lattice design

To meet $\xi_x \ll \nu_z$ required by coherent X-Z instability, simply optimizing the longitudinal dynamics is not sufficient. Therefore, $\beta_x^*$ is reduced from the initial 90 mm to 40 mm to decrease $\xi_x$, which significantly reduces the dynamic aperture and limits the lifetime. Thus, further iterative optimization will be necessary in the future.

### Beam-beam effects

Beam-beam simulations indicate that by carefully selecting a suitable working point, the present parameter design can effectively mitigate coherent X-Z instability. Additionally, the simulated steady-state luminosity essentially meets the design objective.

### Collective effects

PWD-induced bunch lengthening and LWMI are estimated under the assumption of $\text{Im}(Z_\parallel/n)_{\text{eff}} = 0.1\ \Omega$ that is a typical value for modern storage rings. The results indicate that the stretched bunch length is 12 mm, and the threshold bunch current for LWMI is about 4.8 mA, higher than the design bunch current of 3.9 mA.

The transverse effective impedance, approximately calculated as $\text{Im}Z_\perp^{\text{eff}} = 2R/b^2 \cdot \text{Im}(Z_\parallel/n)_{\text{eff}}$ (the Panofsky-Wenzel theorem), is estimated to be 14 kΩ/m with the vacuum chamber half-aperture of $b = 37.5$ mm. Accordingly, the threshold bunch current for TMCI is about 28.9 mA. Therefore, ordinary vacuum chamber impedance will not cause TMCI. However, further study is needed to determine if the addition of special components, especially collimators, will induce TMCI.

The assessment of coupled bunch instabilities driven by higher-order modes of the accelerating cavities and resistive-wall impedance is ongoing.

## SUMMARY

This paper introduces an optimization model of longitudinal dynamics design, which comprehensively considers the coherent X-Z instability, transverse dynamics, and collective effects. It provides crucial support for the physical design of STCF. Further iterative optimization is still needed, particularly in evaluating collective instabilities, including fast ion instabilities and electron cloud effects.